\documentclass[twocolumn,showpacs,preprintnumbers,amsmath,amssymb]{revtex4}

\usepackage{graphicx}
\usepackage{dcolumn}
\usepackage{bm}

\begin{document}


\title{A magneto-electric quantum wheel.}
\author{Alexander Feigel\footnote{Electronic address: sasha@soreq.gov.il}}
\affiliation{%
Soreq NRC,\\
Yavne 81800, Israel}


\date{\today}

\begin{abstract}
 Here we show that self-propulsion in quantum vacuum may be achieved by rotating or aggregating magneto-electric nano-particles. The back-action follows from changes in momentum of electro-magnetic zero-point fluctuations, generated in magneto-electric materials. This effect may provide new tools for investigation of the quantum nature of our world. It might also serve in the future as a ``quantum wheel'' to correct satellite orientation in space.
\end{abstract}

\maketitle

Despite some initial claims of negligible or even zero momentum transfer\cite{Tiggelen2004,Obukhov2008}, recent theoretical studies concur that material objects may acquire momentum from quantum vacuum\cite{Feigel2004,Tiggelen2006,Birkeland2007,Shen2008,Tiggelen2008,Tiggelen2009}. This can be explained qualitatively by considering quantum vacuum as a random fluctuating electromagnetic field (so called zero fluctuations) composed of propagating modes. Each mode possesses both energy and  momentum, causing Lamb splitting of spectral levels \cite{Lamb1947,Bethe1947}, Casimir attraction of objects in an empty space\cite{Casimir1948,Lamoreaux1997,Bressi2002} and other mechanical interactions from nano to astrophysical scales\cite{Davies2005,Capasso2007,Dupays2008}. The total momentum vanishes if the counter-propagating modes cancel mutually. It occurs in all materials except magneto-electrics, which lack both space and time symmetries\cite{Fiebig2005,Eerenstein2006,Arima2008}.

Self-propulsion requires mechanical back-action from an external medium such as ground, water, air or even a quantum liquid\cite{Dreyfus2005,Avron2006}. This can be provided by wheels or propellor-like devices. Mechanical interaction with electro-magnetic radiation can also serve as a mean for moving matter on macro- and nano-scales\cite{Arcizet2006,Eichenfield2009,Groblacher2009,Hugel2002,Kippenberg2008}. It seems that self-propulsion in vacuum, however, can be achieved only by a rocket-like disposal of mass, at least in foreseeable future\cite{Millis2009}.

In this article we demonstrate that aggregating or rotating magneto-electric particles change the momentum of quantum vacuum and, as a consequence they acquire the resulting difference. It follows from momentum conservation: any change in momentum of zero fluctuations is compensated by a corresponding change in the momentum of a material object or electromagnetic field. These new occurrences of the vacuum momentum transfer do not require external means, such as previously proposed modification of the magneto-electric constant by applying external electric and magnetic fields\cite{Feigel2004,Tiggelen2009} or suppressing the quantum vacuum modes by cavity-imposed boundary conditions\cite{Tiggelen2006,Birkeland2007}. 

Consider an object in crossed electric $E_{x}$ and magnetic $B_{y}$ fields (respectively in $x$ and $y$ directions). It will experience a Lorentz force $F$ given by:
\begin{eqnarray}
  F\propto B_{y}\frac{\partial P_{x}}{{\partial t}}, \label{lf1}
\end{eqnarray}
where $P$ is its average electric polarisation. For the sake of simplicity, $P$ will include only the linear dielectric $\epsilon$ and magneto-electric $\chi$ terms: 
\begin{eqnarray}
  P_{x}\propto \epsilon E_{x}+\chi_{xy}B_{y},\label{pol1}
\end{eqnarray}
Merging eqs. (\ref{lf1}) and (\ref{pol1}) results in:  
\begin{eqnarray}
  F\propto B_{y}\frac{\partial \epsilon E_{x}}{{\partial t}}+\chi_{xy}\frac{1}{2}\frac{\partial B^{2}_{y}}{{\partial t}}+B^{2}_{y}\frac{\partial \chi_{xy}}{{\partial t}}.\label{lf2}
\end{eqnarray}
The first term describes a linear dielectric in external fields and includes no quantum vacuum contribution. The second and the third terms describe both classical and quantum phenomena, since $<B^{2}>$ does not vanish due to zero fluctuations of electro-magnetic field.

According to eq. (\ref{lf2}), momentum can be extracted from quantum vacuum by four different approaches. One way is to suppress zero fluctuations by applying some dynamic boundary conditions (by a shrinking cavity for instance) which eliminate the low frequency modes of quantum vacuum, modifying in this way the $<B^{2}_{vac}>$ value\cite{Tiggelen2006,Birkeland2007}. In this case $F$ will be given by:
\begin{eqnarray}
  F_{vac}\propto \chi_{xy}\frac{1}{2}\frac{\partial <B^{2}_{vac}>}{{\partial t}}.\label{lf3}
\end{eqnarray}
Another way is based on achieving a non-zero value for the time derivative of $\chi_{xy}$, resulting in:
\begin{eqnarray}
  F_{vac}\propto <B^{2}_{vac}>\frac{\partial \chi_{xy}}{{\partial t}}.\label{lf4}
\end{eqnarray}
The magneto-electric constant $\chi_{xy}$ consists of intrinsic and induced terms:
\begin{eqnarray}
  \chi_{xy} = \chi^{0}_{xy}+\kappa_{1}E_{x}B_{y}+\kappa_{2}E_{x}+\kappa_{3}B_{y},\label{chi1}
\end{eqnarray}
where $\chi^{0}_{xy}$ is the intrinsic magneto-electricity, while $\kappa_{1}$, $\kappa_{2}$ and $\kappa_{3}$ are the responses to external fields. This leads to a second approach by using time dependant external electric and magnetic fields\cite{Feigel2004,Tiggelen2009}.

The third, previously unforeseen, method to extract momentum is based on rotation of the electro-magnetic material. The magneto-electric constant is a tensor and its value may be changed, therefore, by rotation. The corresponding force is:
\begin{eqnarray}
  F_{vac}\propto <B^{2}_{vac}>\frac{\partial \chi^{0}_{xy}}{{\partial t}}.\label{lf5}
\end{eqnarray}
For instance, rotation by $\pi$ around the $x$ axis changes the sign of $\chi^{0}_{xy}$. In this case, the maximum momentum transfer is obtained by maximum change of the magneto-electric constant $\propto2\chi^{0}_{xy}$.

The total velocity gain of a magneto-electric particle rotated by $\pi$ in quantum vacuum is:
\begin{eqnarray}
  \Delta v_{rot} \approx A\frac{\hbar}{\rho}\frac{2\chi^{0}_{xy}}{a^{4}}\approx A\frac{\hbar}{m}\frac{2\chi^{0}_{xy}}{a},\label{v1}
\end{eqnarray}
where $A\approx10^{-2}$ is a numerical constant, $\rho$ is the density, $a$ is the size and $m\propto\rho a^{3}$ is the mass of the particle. The size defines the highest frequencies $\omega_{cut}$ of quantum vacuum contributing to the particle's momentum\cite{Feigel2004}. The modes with wavelength $\lambda$ ($\lambda = c/\omega$, where $c$ is velocity of light) which are smaller than $a$ do not contribute to the net polarization $P_{x}$ of the object and, therefore, vanish in (\ref{lf1}). The same expression follows from a change in momentum of quantum vacuum\begin{eqnarray}
  p_{vac}\approx A\hbar\chi/a,\label{p1}
\end{eqnarray}
derived by the rigorous renormalization formalism\cite{Birkeland2007,Tiggelen2008,Tiggelen2009}.

Another method for momentum transfer may be based on separation or aggregation of the magneto-electric particles while preserving their orientation. Momentum of quantum vacuum in $N$ magneto-electrics of size $a$ is greater than in a single object of size $L=N^{1/3}a$ (see (\ref{p1}). The difference in momenta corresponds to the velocities:
\begin{eqnarray}
  \Delta v_{agg} \propto A\frac{\hbar}{\rho}\chi^{0}_{xy}\left (\frac{1}{a^{4}}-\frac{1}{L^{4}}\right ).\label{v3}
\end{eqnarray}
Aggregation of the particles affects the boundary conditions of the high-frequency vacuum modes generating greater forces than a shrinking external cavity, which modifies only the low-frequency ones. 

An experimental investigation of momentum transfer in quantum vacuum would be preferably based on rotation or aggregation of magneto-electric materials. In these configurations the contribution of quantum vacuum can be easily isolated, while with alternating electric or magnetic fields\cite{Feigel2004,Tiggelen2009} the quantum and classical forces are difficult to distinguish, see (\ref{lf2}). 

The discussed effect does not interfere with other quantum effects at nano-scales. Quantum fluctuations of the position or of the magneto-electric constant of particles do not affect the average value of their momentum, as a consequence of the conservation of momentum law.

A propulsion engine may be designed by using for instance an addressable array of small magneto-electric particles or wires. Rotating (see Fig. 1) or aggregating (see Fig. 2) these particles will result in velocity:
\begin{eqnarray}
  \Delta V \propto \Delta v\frac{m}{M},\label{v2}
\end{eqnarray}
where $\Delta v$ is given either by (\ref{v1}) or by (\ref{v3}), $m$ and $M$ are the masses of the rotated particles and the entire load correspondingly. Although the proposed engine will consume energy for manipulation of the particles, the propulsion will occur without any loss of mass.

To correct the attitude of a satellite, several degrees per day are required\cite{MontenbruckGill200509}. This corresponds to a tangential velocity of $1\mu m/sec$ at $1m$ radius. This value may be obtained with a magneto-electric constant of $\chi^{0}_{me}\approx 10^{-3}$ and $\approx 1nm$ particles, assuming that roughly half of the satellite mass is composed of magneto-electric material (see eqs. (\ref{v1}) and (\ref{v2}) taking $\rho\approx1g/cm^{2}$). Materials with magneto-electric constants up to $10^{-4}$ at nanometric scale have been already reported, for instance $FeGaO_{3}$ in a weak ($\approx 100 Oe$) external magnetic field\cite{Kubota2004}. One may reasonably assume that this gap will be closed in the future due to investigations in new composites\cite{Arima2008} and advances in micro-patterning techniques\cite{Kida2006}.

To conclude, mechanical action of quantum vacuum on magneto-electric objects may be observable and have a significant value. Rotation or self-assembly of the nano-particles is enough to generate a back-action from zero electro-magnetic fluctuations. The amount of momentum that can be extracted from quantum vacuum by this effect, may have in the future practical implications, depending on advances in magneto-electric materials.


\bibliographystyle{naturemag}

\begin{widetext}
\clearpage
\begin{figure}
\begin{center}
\resizebox{0.5\textwidth}{!}{\includegraphics{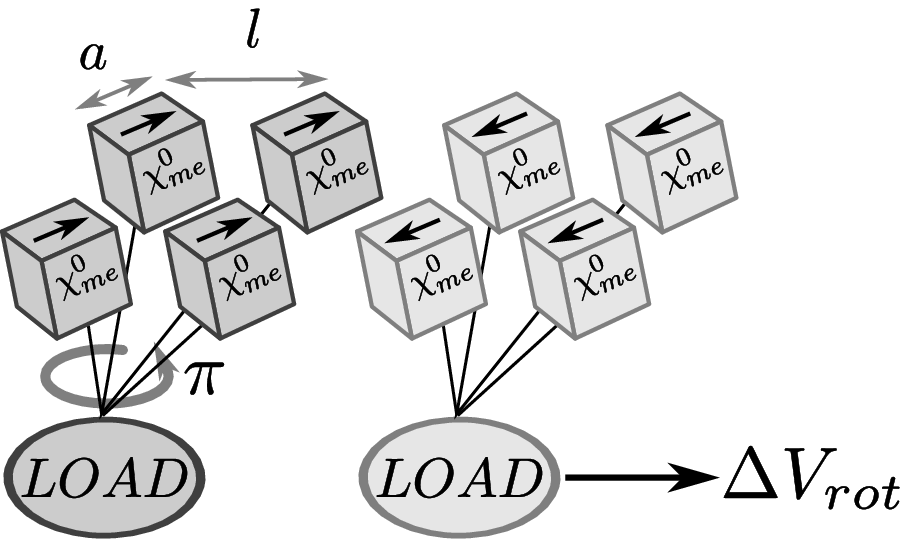}}
\caption{{\bf Propulsion in quantum vacuum by rotation of magneto-electric particles.} Rotation of magneto-electric particles will modify the momentum of quantum vacuum, and generate a corresponding back-action. The resulting velocity is proportional to intrinsic value of the magneto-electric tensor $\chi^{0}_{me}$.}
\label{fig1}
\end{center}
\end{figure}
\begin{figure}
\begin{center}
\resizebox{0.5\textwidth}{!}{\includegraphics{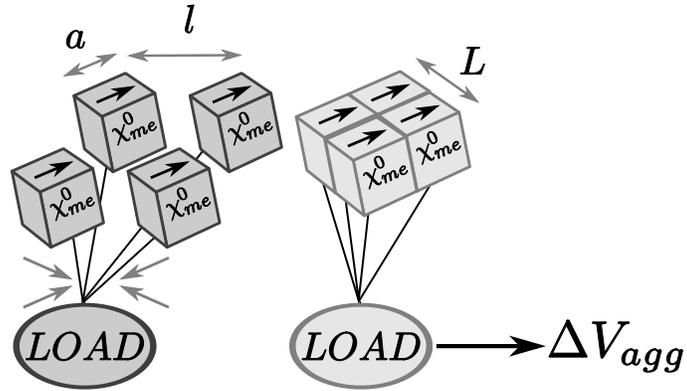}}
\caption{{\bf Propulsion in quantum vacuum by aggregation of magneto-electric particles.} Changes in dimensions of an object with intrinsic $\chi^{0}_{me}$ will affect the momentum of quantum vacuum, generating a back-action. The corresponding velocity increases strongly with smaller particle sizes.}
\label{fig2}
\end{center}
\end{figure}

\end{widetext}
\end{document}